\documentstyle[12pt,epsf]{article}
\input{psfig.tex}
\begin{document}
\begin{titlepage}
\Large
\center{CMB POLARIZATION EXPERIMENTS}
\vspace{0.5cm}
\normalsize
\center{Matias Zaldarriaga\footnote{Email address:
matiasz@arcturus.mit.edu}} 
\vspace{0.3cm}
\center{Department of Physics, MIT, Cambridge, MA 02139 USA}
\vspace{0.3cm}

\begin{abstract}
We discuss the analysis of polarization experiments with particular
emphasis on those that measure the Stokes parameters on a ring on the
sky. We discuss the ability of these experiments to separate the $E$
and $B$ contributions to the polarization signal. The experiment
being developed at Wisconsin
university is  studied in detail, it
will be sensitive to both Stokes
parameters and will concentrate on large scale polarization, scanning a
$43^o$ degree ring. We will also consider another example, 
an experiment that 
measures one of the Stokes parameters
in a $1^o$ ring.
We find that the small ring experiment will be able to
detect cosmological polarization for some models consistent with the
current temperature anisotropy data, for reasonable integration times.
In most cosmological models
large scale polarization is too small 
to be detected by the Wisconsin experiment, but because
both $Q$ and $U$ are measured, separate constraints can be set
on $E$ and $B$ polarization. 
\end{abstract}
\end{titlepage}
\newpage
\section{Introduction}

Temperature anisotropies in the cosmic microwave background (CMB)
can be considered one of the best probes of the early universe. The
higher precision measurements that will become available in the near
future could potentially lead to a precise determination  of a large
number of cosmological parameters \cite{parus,parameters}. 
But the temperature anisotropies are not the only source of information, 
Thomson scattering of the CMB photons will lead to a small degree of 
linear polarization at a level detectable by the future generation of
experiments.

The detection of 
polarization will provide information about our universe that is not
available in the temperature data. For example it can provide an accurate
determination of the ionization history of the universe \cite{zalreio}.
Gravitational waves and vector modes also leave a specific signature in the
polarization \cite{spinlong,kks}. The correlation function of
the Stokes parameters $Q$ and $U$ 
can be used to test the causal structure of our
Universe and thus provides a direct test of inflation \cite{sperzal}. 
The polarization power spectra produced by the current family of 
topological defect
models are also very different from that produced by inflationary models
\cite{selpentur}.

The two future satellite missions, MAP and PLANCK, will measure polarization
as well as temperature anisotropies. The extra information in 
polarization will help improve the constraints these missions can
put on cosmological parameters \cite{parus}. For most
parameters the expected  error bars decrease by a factor of two
and for those related to the reionization history of the universe
the accuracy can be several times better. 

Several ground based polarization experiments are now also under way. 
Some of these experiments plan
to measure the Stokes parameters in pixels on a ring around the north
celestial pole, which simplifies the pointing of the telescope.
The properties of  
the experiment being developed at Wisconsin University \cite{wisc}
are discussed in detail. This experiment will concentrate 
on large angular scale polarization measuring
both $Q$ and $U$ in a $43^o$ ring. We also consider  
another example, an experiment that  measures
only one Stokes parameter in a $1^o$ ring. 

Recent theoretical developments \cite{spinlong,kks} have shown that
rather than describing polarization in terms of $Q$ and $U$ it is
more natural to introduce two scalar quantities $E$ and $B$. The
correlation functions of this variables is what theories most easily
predict and the treatment in terms of $E$ and $B$
takes full account of the spin $2$ nature of polarization. 

The ring like polarization experiments appear as an ideal testing  ground 
for this new formalism  because they are being performed right
now and  also because their simple geometry will enable us to study them
in full detail and build intuition on this new polarization formalism.
We will study for example how this kind of experiments can separate the two
components of polarization $E$ and $B$. Our treatment will also permit
us to explore how changes of the experimental set up, like a change
of the ring size, affect the experiments ability to detect cosmological
polarization.

The outline of the paper is the following: in \S 2 we review previous
results on polarization and describe the way in which general
polarization experiments can be analyzed. In \S 3 we discuss specific
properties of the ring like experiments and in \S 4 we study 
parameter estimation issues in the context of this kind 
of experiments. We make a summary and discuss our results in \S
5.

\section{Correlation Functions}

The CMB radiation field is described by a $2\, \times \, 2$ 
intensity tensor
$I_{ij}$
\cite{chandra}. The Stokes parameters $Q$ and $U$ are defined as 
$Q=(I_{11}-I_{22})/4$ and $U=I_{12}/2$, while the temperature anisotropy 
is given by $T=(I_{11}+I_{22})/4$. The fourth Stokes parameter $V$ that
describes circular polarization is not necessary in standard cosmological 
models because it cannot be generated through the process of Thomson 
scattering. While the temperature is a scalar quantity $Q$ and $U$ are
not. They depend on the direction of observation $\hat n$
and on the two axis $(\hat e_1, \hat e_2)$ 
perpendicular to $\hat n$ used to define them. If for a given 
$\hat n$ the axes $(\hat e_1, \hat e_2)$ are rotated by an angle
$\psi$ such that 
${\hat e_1}^{\prime}=\cos \psi \ {\hat e_1}+\sin\psi \ {\hat e_2}$ 
and ${\hat e_2}^{\prime}=-\sin \psi \ {\hat e_1}+\cos\psi \ {\hat e_2}$
the Stokes parameters change as
\begin{eqnarray}
Q^{\prime}&=&\cos 2\psi \  Q + \sin 2\psi \ U \nonumber \\  
U^{\prime}&=&-\sin 2\psi \ Q + \cos 2\psi \ U
\label{QUtrans} 
\end{eqnarray}

To analize the CMB temperature on the sky it is natural to
expand it in spherical harmonics. These are not appropriate 
for polarization, because   
the two combinations $Q\pm iU$ are quantities of spin $\pm 2$
\cite{goldberg}. They 
should be expanded in spin-weighted harmonics $\, _{\pm2}Y_l^m$ 
\cite{spinlong},
\begin{eqnarray}
T(\hat n)&=&\sum_{lm} a_{T,lm} Y_{lm}(\hat n) \nonumber \\
(Q+iU)(\hat{n})&=&\sum_{lm} 
a_{2,lm}\;_2Y_{lm}(\hat{n}) \nonumber \\
(Q-iU)(\hat{n})&=&\sum_{lm}
a_{-2,lm}\;_{-2}Y_{lm}(\hat{n}).
\label{Pexpansion}
\end{eqnarray}
To perform this expansion, $Q$ and $U$ in equation (\ref{Pexpansion})
are measured relative to $(\hat e_1, \hat e_2)=(\hat e_\theta, \hat
e_\phi)$, the unit vectors of the spherical coordinate system.
There is an equivalent expansion using tensors on the
sphere \cite{kks}.
The coefficients $_{\pm 2}a_{lm}$ 
are observable on the sky and their power spectra
can be 
predicted for different cosmological models. Instead of $_{\pm 2}a_{lm}$
it is convenient
to use their linear combinations
$a_{E,lm}=-(a_{2,lm}+a_{-2,lm})/2$ and 
$a_{B,lm}=-(a_{2,lm}-a_{-2,lm})/2i$, which transform differently 
under parity.
Four power spectra are needed to
characterize fluctuations in a gaussian theory,
the autocorrelation between 
$T$, $E$ and $B$ and the cross correlation of $E$ and $T$.
Because of parity considerations the cross-correlations
between $B$ and the
other quantities vanish and one is left with
\begin{eqnarray}
\langle a_{X,lm}^{*}
a_{X,lm^\prime}\rangle &=& \delta_{m,m^\prime}C_{Xl}
\nonumber \\
\langle a_{T,lm}^{*}a_{E,lm}\rangle&=&\delta_{m,m^\prime}C_{Cl}
\label{Cls},
\end{eqnarray}
where
$X$ stands for $T$, $E$ or $B$, $\langle\cdots \rangle$
means ensemble average and $\delta_{i,j}$ is the Kronecker delta.  

For the purpose of this work it is more useful to rewrite 
equation (\ref{Pexpansion}) as
\begin{eqnarray}
T(\hat{n})&=&\sum_{lm} a_{T,lm} Y_{lm}(\hat n) \nonumber \\
Q(\hat{n})&=&-\sum_{lm} 
a_{E,lm} X_{1,lm} 
+i a_{B,lm}X_{2,lm} \nonumber \\
U(\hat{n})&=&-\sum_{lm} 
a_{B,lm} X_{1,lm}-i a_{E,lm} X_{2,lm}
\label{Pexpansion2}
\end{eqnarray}
where we have introduced
$X_{1,lm}(\hat{n})=(\;_2Y_{lm}+\;_{-2}Y_{lm})/2$
and $X_{2,lm}(\hat{n})=(\;_2Y_{lm}-\;_{-2}Y_{lm})/ 2$.
They satisfy $X^{*}_{1,lm}=X_{1,l-m}$  and $X^*_{2,lm}=-X_{2,l-m}$ which
together with $a_{E,lm}=a_{E,l-m}^*$ and
$a_{B,lm}=a_{B,l-m}^*$ make  $Q$ and $U$ real.

In fact $X_{1,lm}(\hat{n})$ and $X_{2,lm}(\hat{n})$ have the form,
$X_{1,lm}(\hat{n})=\sqrt{(2l+1) / 4\pi}$ $F_{1,lm}(\theta)\ e^{im\phi}$
and $X_{2,lm}(\hat{n})=\sqrt{(2l+1) / 4\pi}$ $F_{2,lm}(\theta)\ e^{im\phi}$, 
$F_{(1,2),lm}(\theta)$ can be calculated in terms of Legendre
polynomials \cite{kks} \footnote{A subroutine that calculates this
functions is available at http://arcturus.mit.edu/\~{}matiasz/CMBFAST}:
\begin{eqnarray}
F_{1,lm}(\theta)&=&2 \sqrt{(l-2)!(l-m)! \over (l+2)!(l+m)!}
[ -({l-m^2 \over \sin^2\theta} 
+{1 \over 2}l(l-1))P_l^m(\cos \theta)
\nonumber \\
&&+(l+m) {\cos \theta \over \sin^2 \theta} 
P_{l-1}^m(\cos\theta)] \nonumber \\
F_{2,lm}(\theta)&=&2 \sqrt{(l-2)!(l-m)! \over (l+2)!(l+m)!}{m \over
\sin^2 \theta}
[ -(l-1)\cos \theta P_l^m(\cos \theta)\nonumber \\
&&+(l+m) P_{l-1}^m(\cos\theta)]. 
\end{eqnarray}
Note that $F_{2,lm}(\theta)=0$ if $m=0$, as it  must to make the
Stokes parameters real.

The correlation functions can be calculated using equations (\ref{Cls}) 
and (\ref{Pexpansion2}), 
\begin{eqnarray}
\langle T(1)T(2) \rangle&=&\sum_l C_{Tl} [\sum_m \;_0Y_{lm}^*(1) 
\;_0Y_{lm}(2)] \nonumber \\ 
\langle Q(1)Q(2) \rangle&=&\sum_l C_{El} [\sum_m X_{1,lm}^*(1) 
X_{1,lm}(2)]+C_{Bl}[\sum_m X_{2,lm}^*(1) 
X_{2,lm}(2)] \nonumber \\ 
\langle U(1)U(2) \rangle&=&\sum_l C_{El} [\sum_m X_{2,lm}^*(1) 
X_{2,lm}(2)]+C_{Bl}[\sum_m X_{1,lm}^*(1) 
X_{1,lm}(2)] \nonumber \\ 
\langle T(1)Q(2) \rangle&=&\sum_l C_{Cl} [\sum_m \;_0Y_{lm}^*(1) 
X_{1,lm}(2)] \nonumber \\
\langle T(1)U(2) \rangle&=&i\sum_l C_{Cl} [\sum_m \;_0Y_{lm}^*(1) 
X_{2,lm}(2)] 
\label{correl}
\end{eqnarray}
where $1$ and $2$ stand for the two directions in the sky
$\hat{n}_1$ and  $\hat{n}_2$.
These expressions can be further simplified
using the addition theorem for the spin harmonics \cite{primer},
\begin{equation}
\sum_m \;_{s_1} Y_{lm}^*(\hat{n}_1) 
\;_{s_2} Y_{lm}(\hat{n}_2)=\sqrt{2l+1 \over 4 \pi} 
\;_{s_2} Y_{l-s_1}(\beta,\psi_1)e^{-is_2\psi_2}
\label{addtheo}
\end{equation}
where $\beta$ is the angle between $\hat{n}_1$ and
$\hat{n}_2$, and $(\psi_1$,$\psi_2)$ are the angles 
$(\hat e_\theta, \hat e_\phi)$ at $\hat{n}_1$ and
$\hat{n}_2$ need to be rotated to become aligned with
the great circle going through both points.
In the case of the temperature equation(\ref{addtheo}) gives the usual
relation, 
\begin{equation}
\langle T_1T_2 \rangle=\sum_l {2l+1 \over 4 \pi}
C_{Tl} P_l(\cos \beta).
\end{equation}

For polarization the addition
relations for $X_{1,lm}$ and $X_{2,lm}$ are calculated from equation 
(\ref{addtheo}),
\begin{eqnarray}
\sum_m X_{1,lm}^*(1) 
X_{1,lm}(2)&=&{2l+1 \over 4 \pi} 
[F_{1,l2}(\beta)  \cos 2\psi_1 \cos 2\psi_2 - F_{2,l2}(\beta)
\sin 2\psi_1 \sin 2\psi_2] \nonumber \\
\sum_m X_{2,lm}^*(1) 
X_{2,lm}(2)&=&{2l+1 \over 4 \pi} 
[F_{1,l2}(\beta)  \sin 2\psi_1 \sin 2\psi_2 - F_{2,l2}(\beta)
\cos 2\psi_1 \cos 2\psi_2] \nonumber \\
\sum_m X_{1,lm}^*(1) 
X_{2,lm}(2)&=&i{2l+1 \over 4 \pi} 
[F_{1,l2}(\beta)  \sin 2\psi_1 \cos 2\psi_2 + F_{2,l2}(\beta)
\cos 2\psi_1 \sin 2\psi_2] \nonumber \\ 
\sum_m \;_0Y_{lm}^*(1) 
X_{1,lm}(2)&=&{2l+1 \over 4 \pi} 
F_{1,l0}(\beta)  \cos 2\psi_2 
\nonumber \\
\sum_m \;_0Y_{lm}^*(1) 
X_{1,lm}(2)&=&-i{2l+1 \over 4 \pi} 
F_{1,l0}(\beta)  \sin 2\psi_2 
\label{addrelpol}
\end{eqnarray}
where we can equivalently write $F_{1,l0}(\beta)=\sqrt{(l-2)!/(l+2)!}
P_l^2(\beta)$.

The correlations in 
equation (\ref{correl}) with the sums given by equation 
(\ref{addrelpol}) are all what is 
needed to analize any given experiment. These
relations are simple to understand, as pointed out in \cite{kks} the
natural coordinate system to express the correlations is one in which
both $(\hat e_1, \hat e_2)$ reference frames 
are chosen to be aligned with the great circle
connecting the two directions (1 and 2); in that case we have \cite{kks}
\begin{eqnarray}
\langle Q_{r}(1)Q_{r}(2) \rangle&=&\sum_l {2l+1 \over 4 \pi} [C_{El}
F_{1,l2}(\beta)-C_{Bl} F_{2,l2}(\beta)]  \nonumber \\ 
\langle U_{r}(1)U_{r}(2) \rangle&=&\sum_l {2l+1 \over 4 \pi}
[C_{Bl} F_{1,l2}(\beta)-C_{El} F_{2,l2}(\beta) ] \nonumber \\ 
\langle T(1)Q_{r}(2) 
\rangle&=&\sum_l {2l+1 \over 4 \pi} C_{Cl} F_{1,l0}(\beta)\nonumber \\
\langle T(1)U_{r}(2) \rangle&=&0
\label{QUr}
\end{eqnarray}
the subscript $r$ 
here indicate that the Stokes parameters are measured in this
particular coordinate system.
We can use
the transformation laws in equation (\ref{QUtrans})
to write $(Q,U)$ in terms of $(Q_r,U_r)$ and
then using equation (\ref{QUr}) for their correlations one can recover
our final result (given by equations
(\ref{correl}) and (\ref{addrelpol})).

When analyzing an experiment 
we can arrange the measured values of $T$, $Q$ and $U$ in a vector 
${\bf x}=(T_1,Q_1,U_1,...T_i,Q_i,U_i...)$, the subscript
labels the pixel. 
The measured Stokes parameters  at each pixel
have two contributions, the first coming from
the cosmological signal we are interested in measuring,
given by equation (\ref{Pexpansion2})  and the second 
from the noise in the detectors. The correlation matrix of ${\bf x}$ is 
$\langle {\bf x}_i{\bf x}_j\rangle \equiv
{\bf C}_{ij}={\bf S}_{ij}+{\bf N}_{ij}$, where the signal
correlation matrix ${\bf S}$ is given in equations (\ref{correl}) and
(\ref{addrelpol}) and ${\bf N}$ is the correlation matrix of the
noise. With this correlation matrix the full likelihood can be
calculated,
\begin{equation}
L({\bf x}
|C_{Xl})\propto {1 \over \sqrt{\det C}} \exp[-{1\over 2} {\bf
x}^{T} {\bf C}^{-1} {\bf x}],
\end{equation}
here $C_{Xl}$ stands for the complete set of power spectra that
describe the theory under consideration, 
$C_{Xl}=\{C_{Tl},C_{El},C_{Bl},C_{Cl}\}$.
If the set of power spectra are given in terms of a set of parameters,
the likelihood can be maximized to find 
parameters that best fit the data.

\section{Ring Experiments}

There are several ground based 
polarization experiments now under way. Some of
these experiments plan to measure the Stokes parameters $Q$ and $U$
in pixels that form a ring on the sky.  The simple geometry of
the sky patch is perfect for understanding the relation between $E$
and $B$ polarization and the Stokes parameters.  
It is the aim of this section to analyze this kind of experiments.

Rather than doing the analysis in terms of correlations in real space
as was 
suggested in the previous section, in
this case it is simple to diagonalize the signal correlation matrix.
If the 
noise is uncorrelated and equal from pixel to pixel 
(ie. the noise correlation matrix is
proportional to the identity matrix) we simultaneously diagonalize
the noise and signal correlation matrices. This will allow us
to make a more detailed analysis of the characteristics of this type
of experiments. 

Let us assume that the experiment measures the Stokes
parameters in a number ($N_{pix}$) of pixels on a ring of radius
$\theta$ with a gaussian beam of width $\theta_{fwhm}$.
We will take the noise correlation matrix to be
diagonal, $\langle {N}_{(Q,U)i} {N}_{(Q,U)j}\rangle = \sigma_P^2
\delta_{ij}$ and uncorrelated between $Q$ and $U$,
${N}_{(Q,U)j}$ are the noise contribution to the measured $Q$ and $U$
in pixel $j$.
If $\phi_j=2\pi(j-1)/N_{pix}$ denotes the angle along the ring, the Stokes
parameters in each pixel are given by (equation (\ref{Pexpansion2}) 
plus the noise
contribution) 
\begin{eqnarray}
Q_j&=&-\sum_{lm} \sqrt{(2l+1) / 4\pi}B_{lm}[ 
a_{E,lm}  F_{1,lm}(\theta) \nonumber \\
&&+i a_{B,lm} F_{2,lm}(\theta) ]e^{im\phi_j} 
+ N_{Qj} \nonumber \\
U_j&=&-\sum_{lm} \sqrt{(2l+1) / 4\pi}B_{lm}[ 
a_{B,lm} F_{1,lm}(\theta) \nonumber \\
&&-i a_{E,lm} F_{2,lm}(\theta) 
]e^{im\phi_j}+ N_{Uj}, 
\label{Pexpansion3}
\end{eqnarray}
here $B_{lm}$ encodes the information the beam and scan
pattern of the experiment. 
In an appendix we make the derivation of these functions for
the Wisconsin experiment. In what follows we will take
$B_{lm}^2=\exp{[-l(l+1)\sigma_b^2]}$ with $\sigma_b=
\theta_{fwhm}/2\sqrt{2 \ln 2}$
which is a good first approximation. 

Diagonalizing the signal correlation matrix is very simple and
intuitive: 
if one considers the Fourier transform of the data then each
mode will pick a particular value of $m$ in equation
(\ref{Pexpansion3}). In particular let us consider the transformed data
set
\begin{eqnarray}
{\bar Q}^k &\equiv& {1\over N_{pix}} \sum_j Q_j
e^{-i k \phi_j} \nonumber \\
&=& -\sum_{l\geq |k|} \sqrt{(2l+1) / 4\pi}B_{lk}[ 
a_{E,lk}  F_{lk}^1(\theta) 
+i a_{B,lk} F_{lk}^2(\theta) ]+{\bar N}_Q^k \nonumber \\ 
{\bar U}^k &\equiv& {1\over N_{pix}} \sum_j U_j
 e^{-i k \phi_j} \nonumber \\
&=& -\sum_{l\geq |k|} \sqrt{(2l+1) / 4\pi}B_{lk}[ 
a_{B,lk} F_{lk}^1(\theta)-i a_{E,lk}  F_{lk}^2(\theta) 
]+{\bar N}_U^k 
\label{FFT}
\end{eqnarray}
where $k$ runs from $-N_{pix}/2 \leq k \leq N_{pix}/2$. 
To get to the last expression we have used that,
\begin{eqnarray}
\sum_{j=1}^{N_{pix}} e^{
i(m-k)\phi_j}&=&\sum_{j=0}^{N_{pix}-1}(e^{i(m-k)2\pi/N_Pix})^j
\nonumber \\
&=&(1-e^{i(m-k)2\pi})/(1-e^{i(m-k)2\pi/ N_{pix}}) \nonumber \\
&=& N_{pix} \delta_{m,k+n N_{pix}}
\label{sum1}
\end{eqnarray}
where $n$ is any integer. The sum in equation (\ref{sum1}) is only
different from zero if $k=m$ or if it differs by a multiple of
$N_{pix}$. To obtain equation (\ref{FFT}) we ignored the terms 
that ``leak'' power from higher harmonics, only $k=m$ is considered. 
This is a reasonable
assumption because higher harmonics are suppressed by beam smearing.
In fact we should always choose $N_{pix}$ large enough so as to
oversample the beam and not loose the information in the smaller scales,
this immediately guaranties that the aliased power is negligible.

In equation (\ref{FFT}) 
we have labeled the Fourier transforms of the noise contributions 
${\bar N}_{(Q,U)}^k$ which under the assumption of uniform
uncorrelated noise satisfies, 
\begin{eqnarray}
\langle {\bar N}_Q^{k*}{\bar N}_Q^{k^\prime} \rangle&=&
w_P^{-1} \delta_{k,k^{\prime}} \nonumber \\   
\langle {\bar N}_U^{k*}{\bar N}_U^{k^\prime} \rangle&=&
w_P^{-1} \delta_{k,k^{\prime}} \nonumber \\   
\langle {\bar N}_Q^{k*}{\bar N}_U^{k^\prime} \rangle&=& 0
\nonumber \\   
\langle {\bar N}_U^{k*}{\bar N}_Q^{k^\prime} \rangle&=& 0
\label{noisecorr}
\end{eqnarray}
where $w^{-1}_P\equiv \sigma_P^2 /N_{pix}$. 

We will take
the data set to be the Fourier coefficients 
rather than the Stokes
parameters.
The signal part of each Fourier component receives contributions
only from multipoles with $m=k$, equation  (\ref{Cls}) then implies
that only components with the same value of
$k$ will be correlated. 
The correlation matrix is block diagonal,
\begin{eqnarray}
\langle {\bar Q}^{k*}{\bar Q}^{k^\prime} \rangle&=&
\delta_{k,k^{\prime}} 
\sum_{l\geq k} {(2l+1) / 4\pi}B_{lk}^2[ 
C_{E,l}  F_{1,lk}^2(\theta) 
+C_{B,l} F_{2,lk}^2(\theta) ] \nonumber \\
&& + \delta_{k,k^{\prime}} w_P^{-1} \nonumber \\      
\langle {\bar U}^{k*}{\bar U}^{k^\prime} \rangle&=&
\delta_{k,k^{\prime}}
\sum_{l\geq k} {(2l+1) / 4\pi}B_{lk}^2[ 
C_{E,l}  F_{2,lk}^2(\theta) 
+C_{B,l} F_{1,lk}^2(\theta) ]  \nonumber \\   
&& + \delta_{k,k^{\prime}} w_P^{-1} \nonumber \\      
\langle {\bar U}^{k*}{\bar Q}^{k^\prime} \rangle &=&
\delta_{k,k^{\prime}}  
\sum_{l\geq k} i {(2l+1) / 4\pi}B_{lk}^2 
(C_{E,l}+C_{B,l})  F_{1,lk}(\theta) F_{2,lk}(\theta)
 \nonumber \\   
\langle {\bar Q}^{k*}{\bar U}^{k^\prime} \rangle&=&-
\langle {\bar U}^{k*}{\bar Q}^{k^\prime} \rangle, 
\label{signalcorr}
\end{eqnarray}
``$*$'' means complex conjugate.
An important point is that this transformation does not rely on the
particular shape or other property of the power spectra,  if the CMB 
is a statistically
isotropic and homogeneous random field taking the Fourier transform
will always
make the correlation matrix block diagonal.  
The correlation matrix is hermitian, and the cross correlation
between ${\bar U}^{k}$ and ${\bar Q}^{k^\prime}$ is imaginary. It will
be more convenient to change variables and use $i {\bar U}^{k}
\rightarrow  {\bar U}^{k}$, the data set will then be 
${\bf x}_k=({\bar Q}^k,i{\bar U}^k)$. The correlation matrix for each
$k$ is given by equation (\ref{signalcorr}) but without the $i$ in the
cross term. 

In the Fourier domain the
likelihood becomes a product of independent gaussians, one for each $k$,
\begin{equation}
L({\bf x}|C_{Xl})\propto \prod_k {1 \over 
\sqrt{\det C_k}} \exp[-{1\over 2} {\bf
x}^{\dag}_k {\bf C}_k {\bf x}_k],
\end{equation}
where the ${\bf C}_k$ matrices are $2\times 2$ 
given by equation (\ref{signalcorr})
and $\dag$ means transpose and complex conjugate. 
Clearly if only one of the Stokes parameters is
measured then the likelihood is just the product of several one
dimensional gaussians and if the temperature is also included in the
analysis ${\bf C}_k$ turns into a $3 \times 3$ matrix.
 
Equation (\ref{signalcorr}) shows that if a ring experiment only
measures one of the Stokes parameters, either ${\bar Q}^k$ or ${\bar
U}^k$ it will not be able to separate the $E$ and $B$ contributions in
a model independent way because both contributions enter in the
expressions summed
together. This is similar to what 
happens if one is interested 
in separating the contributions from
density perturbations and gravitational waves
to the temperature
anisotropies. It is only when one
{\it assumes} a shape for the power spectra that one can fit the observed
temperature spectra as a sum of these two separate contributions and
infer the presence of gravity waves. In a similar way if we only
measure ${\bar Q}^k$ and we {\it assume} different shapes for both
$C_{El}$ and $C_{Bl}$ we could determine each contribution. On the
other hand if both Stokes parameters are measured then we could
separate the $E$ and $B$ contribution directly as they enter
differently in each correlation (multiplied by $F_{1,lk}^2(\theta)$ and
$F_{2,lk}^2(\theta)$ in $Q$ and the other way around in $U$).

Another interesting property of equation (\ref{signalcorr}) is that
the $k=0$ $Q$ mode does not receive any contribution from the $B$
channel because $F_{2,l0}(\theta)\equiv 0$, the converse is true for $U$. 
For example a non zero signal in ${\bar
U}^0$ implies the presence of $B$ and thus of gravity waves or vector
modes. Of course this determination suffers from a large cosmic
variance, because we are only measuring {\it one} realization.  

\subsection{Small scale experiment}

Let us now analize how the signal in each mode depends on the power
spectrum. We will do this first for a small ring experiment that 
measures only $Q$ on a ring with $\theta=1^o$, 
$\theta_{fwhm}=0.22^o$. We selected a small ring scanned with a fraction of a
degree resolution to make the experiment sensitive to small scale
polarization. We will choose   
$w^{-1}_P=(0.9\mu K)^2$, corresponding to a
receiver noise of $1.2 mK/\sqrt{Hz}$ and three weeks of observations. 
This numbers are chosen to be representative of what a ground
experiment might achieve.

The correlation matrix is 
\begin{equation}
\langle {\bar Q}^{k*}{\bar Q}^{k^\prime} \rangle=
\delta_{k,k^{\prime}}\sum_{l} {(2l+1) / 4\pi}(
W_{1,lk} C_{E,l}+W_{2,lk} C_{B,l})
+\delta_{k,k^{\prime}}w^{-1}_P. 
\label{corrprinc}
\end{equation} 
Both signal and noise parts are diagonal and 
we have introduced the window functions $W_{1,lk}=B_{lk}^{2}\
F_{1,lk}^2(\theta)$  and
$W_{2,lk}=B_{lk}^{2}\
F_{2,lk}^2(\theta)$ which describe how each Fourier mode depends on the
underlying power spectra. Let us consider what the experiment would
measure if the underlying cosmological model was standard CDM. We will
assume there are no gravity waves or vector modes so we will only
consider $E$ type polarization. We will generalize our  analysis
later.  

\begin{figure}[t]
\vspace*{6cm}
\caption{Window functions for the small ring experiment for several
values of $k$. The SCDM spectrum is also shown for comparison, we
actually plot $(l+1)C_{El}/2\pi$ which is the relevant measure of
power for a  linear $l$ scale (the
normalization is arbitrary).} 
\includegraphics{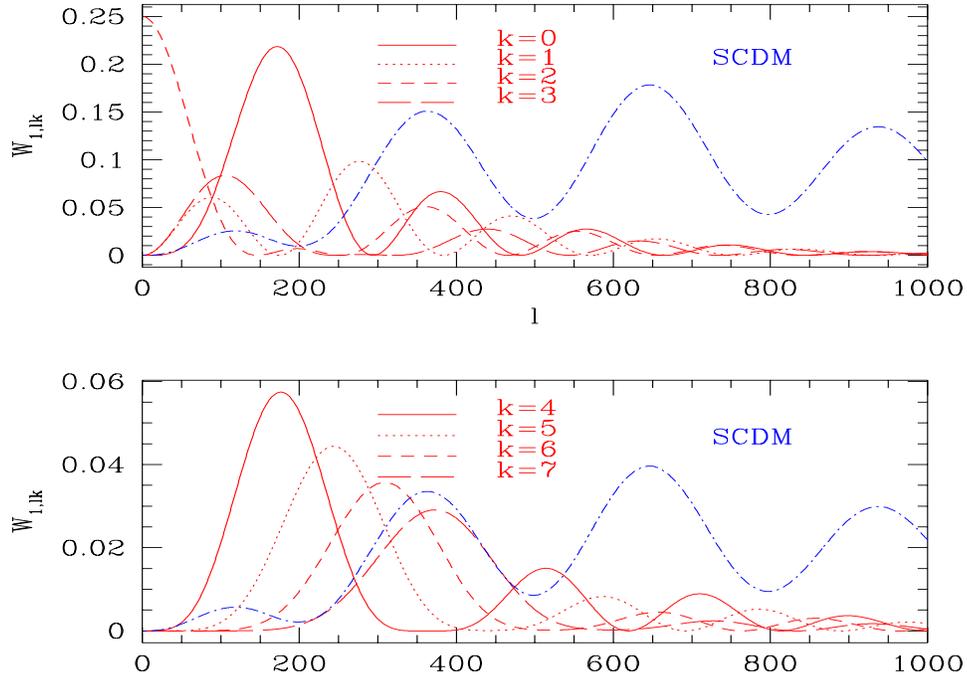}
\label{figure1}
\end{figure}

Figure \ref{figure1} shows the window functions $W_{1,lk}$ for several values of
$k$. Only multipoles $l\geq |k|$ contribute to Fourier component $k$. As
$|k|$ grows the peaks of the window functions move to higher $l$ (we
only show results for the $k >0$ modes but they also apply to the $k<0$
ones). The 
window functions are very broad in $l$ space, because their
width is inversely proportional to the
smallest dimension of the sky patch one is observing. For large
values of $l$ the window functions are cut off  by the effect of the
beam smearing, $B^2_{lk}$. 

Figure \ref{figure1} also shows 
that the most important contribution to the signal in this
experiment comes from the second polarization acoustic peak at $l\sim
350$, the first peak at $l\sim 100$ contributes to a smaller extent. 
Higher peaks are inaccessible because of beam smearing.

An interesting point to note is that the $k=2$ window
function peaks at low $l$ (the lowest $l$ in the polarization
expansion is by definition $l=2$), 
so it is this mode rather than $k=0$ or $1$ that gets
the contribution from the largest angular scale perturbations. 
To understand why let us consider a small patch of the sky 
containing the $\theta=1^o$ ring and 
a polarization field that is
constant over the patch. This polarization field is produced by the
very large angular scale perturbations, the low $l$ modes. Without
loss of generality let us call ${\hat e_x}$ the direction of the 
polarization vectors,
${\hat e_y}$ will be perpendicular to that. 
The Stokes parameters are measured
relative to the $({\hat e_\theta},{\hat e_\phi})$ basis 
which rotates as one moves along the ring.  Although
the polarization amplitude $P$ is constant over the whole patch and the
direction is always ${\hat e_x}$ the Stokes parameters along the ring 
(measured in the $({\hat e_\theta},{\hat e_\phi})$ basis)
are given by 
\begin{eqnarray}
Q_j&=&P \cos 2\phi_j \nonumber \\ 
U_j&=&P \sin 2\phi_j,
\end{eqnarray}
and thus the $k=\pm 2$ Fourier modes will get all the signal.
The coordinate system used to describe the Stokes parameters,
which is rotating as one moves along the ring, 
and their spin 2 nature, which tells us how $Q$ and $U$ transform
under this rotation, makes
the large angular scale perturbations contribute only to the 
$k=\pm 2$ mode. This is another illustration of the spin 2 nature of
polarization. 

\begin{figure}[t]
\vspace*{6cm}
\caption{Window functions for the small ring experiment for 
$k=4,8$ and two ring sizes $\theta=1^o,0.8^o$.}
\includegraphics{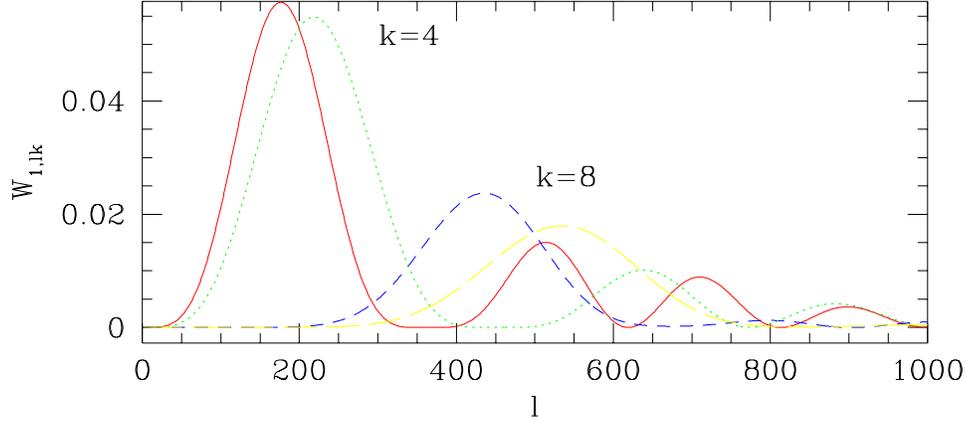}
\label{figure2}
\end{figure}

Another interesting thing to consider is the effect of the size of the
ring on the window functions. The window functions are oscillating
functions of $l$ and changes in the ring size modify the period of these
oscillations. In particular the separation between peaks scales as
$1/ \theta$, and at the same time as the peaks become more
separated they also become broader. This is most easily understood in
the small scale limit. If we can neglect the curvature of the sky,
which is an excellent approximation for this experiment, we can use Fourier
transforms instead of spherical harmonics and write
\cite{urospol,spinlong}
\begin{equation}
Q({\bf \theta})=(2\pi)^{-2}\int d^2l E({\bf l}) \cos 2(\phi-\phi_{
l})\; e^{i l \theta \cos (\phi -\phi_{l})}
\end{equation}
where ${\bf \theta}$ labels the Fourier mode with $(l_x+il_y)=l
\exp{(i\phi_{l})}$. We have again considered only $E$ type
polarization and here $\langle E({\bf l}_1) E({\bf l}_2)
\rangle=C_{El_1} \delta^D({\bf l}_2-{\bf l}_1)$, $\delta^D({\bf
x})$ is the Dirac $\delta-$function. A straightforward
calculation leads to
\begin{equation}
{\bar Q}^k=-{1 \over  2 \pi} \int d^2l E({\bf l}) e^{ -i k \phi_{
l}}\  {1\over 2} [J_{k+2}(u)+J_{k-2}(u)],
\end{equation}
where $u=l\theta$ and $J_n(u)$ are Bessel functions. The correlation
matrix is then given by 
\begin{equation}
\langle {\bar Q}^{k*}{\bar Q}^{k^\prime}\rangle=\delta_{k,k^\prime}
\int {l dl \over 2\pi}\ C_{El}\ {1\over 4}[J_{k+2}(u)+J_{k-2}(u)]^2
\end{equation}
which means that the window function in the small scale limit is 
$W_{1,lk}={1\over 4}[J_{k+2}(u)+J_{k-2}(u)]^2$. The effect of
changing the
size of the ring is straight forward to understand 
in this expression, each $W_{1,lk}$ is
only a function of $u=l\theta$ so changes in $\theta$ just cause a
stretching of the window function. A similar analysis for the other
set of 
window function yields $W_{2,lk}={1\over
4}[J_{k_1+2}(u)-J_{k_1-2}(u)]^2$. Beam smearing only
adds a $B_{lk}^2$ to the expressions of both $W_{(1,2),lk}$ . Figure
\ref{figure2} shows two examples of these window functions for two ring sizes,
the effect of scaling together with the $B_{lk}^2$ factor is clear.

\begin{figure}[t]
\vspace*{6cm}
\caption{Signal to noise in the different modes for the small ring
experiment assuming $w_P^{-1}=(0.9 \mu K)^2$ and COBE normalized SCDM as the
underlying model. Two different ring sizes are plotted, 
in each case the effect of changing the beam width from $0.22^o$(upper
curve) to $0.2^o$ (lower curve) is also shown.}
\includegraphics{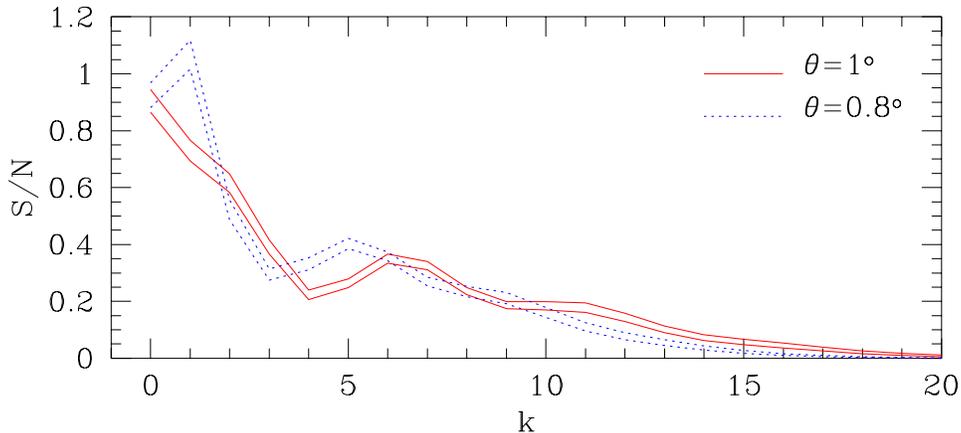}
\label{figure3}
\end{figure}

In figure \ref{figure3}
 we show the S/N for each of the Fourier modes, that is the
ratio of the signal to
the noise part of the correlation 
in (equation \ref{corrprinc}), as a function
of $k$. As expected the signal to noise  decreases with
increasing $k$ because of beam smearing. The effect of decreasing the
ring size is evident in the figure, because the window functions
for successive $k$
become more separated and at the same time broader 
the signal in the experiment gets
concentrated in fewer Fourier modes. 
Also note that there is a peak at $k\sim 7$ when the ring is
$\theta=1^o$  because this window function just matches
the second acoustic peak and is then able to pick more power. The
position of this peak is shifted to lower $k$ when the size of the
ring is decreased.  

It is important to recognize that most modes have a $S/N$ smaller than
one, which will make the polarization hard to detect. Other models
different from SCDM would produce a higher signal, but reducing
the noise in the experiment would crucially   improve its
sensitivity to cosmological polarization. On the other 
hand the overall S/N of the experiment not just mode by mode
is better, opening the
possibility that cosmological polarization will be detected in the
near future with this type of experiments.

\subsection{Wisconsin experiment}

\begin{figure}[t]
\centerline{\psfig{figure=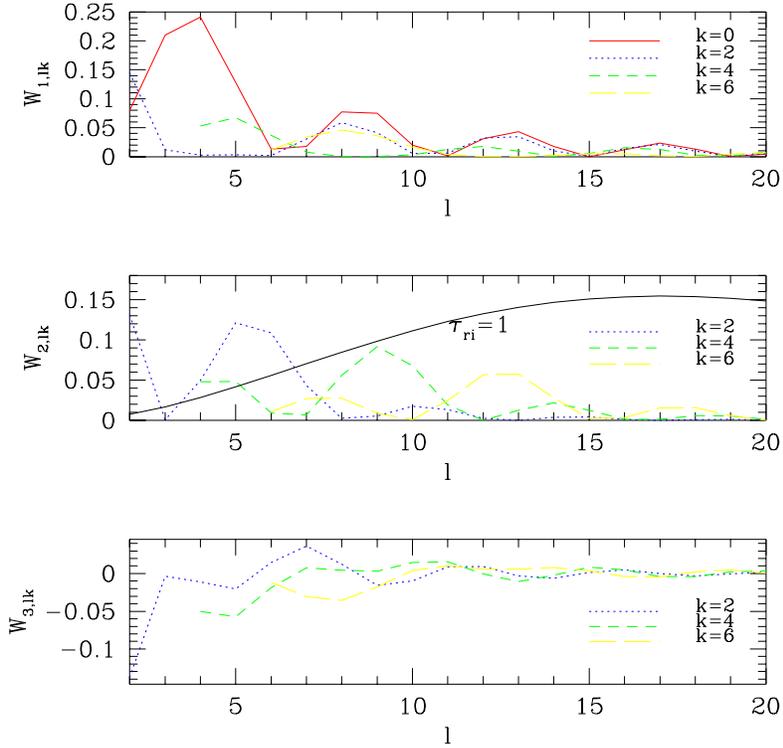,height=4.2in}}
\caption{Window functions for the Wisconsin experiment for several
values of $k$. For comparison in the middle pannel we show the $E$
spectra, $(l+1)C_{El}/2\pi$, for SCDM but with $\tau_{ri}=1$ (the
normalization is arbitrary).}  
\label{figure4}
\end{figure}

Let us now consider the Wisconsin experiment, because it
will actually measure both $Q$ and $U$ on the ring it is in principle
capable of distinguishing between $E$ and $B$
polarizations. The ring will be $\theta=43^o$, scanning the sky on the
vertical at Wisconsin. It will have
$\theta_{fwhm}=7^o$ and
$w^{-1}_P=(0.4 \mu K)^2$, the experiment is thus sensitive to large
scale polarization. 
Unfortunately the large scale polarization is very
small in most cosmological models for the sensitivity of this experiment.

Figure \ref{figure4} shows some of  
the window functions for the experiment with $W_{1,lk}=B_{lk}^2\
F_{1,lk}^2(\theta)$, $W_{2,lk}=B_{lk}^2\ F_{2,lk}^2(\theta)$ 
and $W_{3,lk}=B_{lk}^2 \
F_{1,lk}(\theta)\ F_{2,lk}(\theta)$.
Note that the window
functions for the cross correlation between $Q$ and $U$ 
can actually be negative.
The properties of these window functions are the same as those
discaused above for the small scale experiment, but note that because
both the ring and the beam are much bigger now, 
the experiment is only sensitive to $l\leq 25$.

\begin{figure}[t]
\vspace*{6cm}
\caption{S/N for the Stokes parameters in the Wisconsin experiment.
At low $k$ for both $Q$ and $U$ the lower curve corresponds to COBE
normalized SCDM
with $\tau_{ri}=1.0$, the upper curves are for a model with
$\sigma_E^2=2\mu K^2$ and $\sigma_B^2=0\mu K^2$. } 
\includegraphics{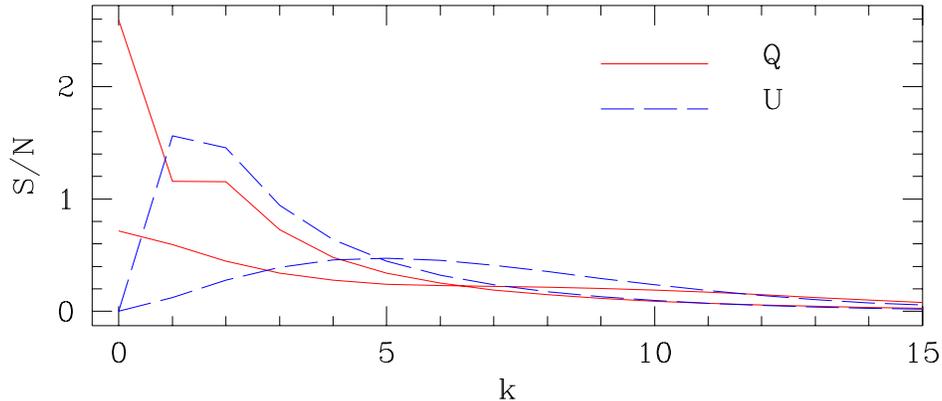}
\label{figure5}
\end{figure}

We will illustrate the features of this experiment by considering
different underlying toy models, for simplicity we will take the
power per logarithmic interval of $E$ and $B$ to be constant,
$l(l+1)C_{(E,B),l}/2\pi=\sigma^2_{(E,B)}$.
Figure \ref{figure5} shows the $Q$ and $U$ $S/N$,
the ratio of the
signal to the noise contributions to the autocorrelations,
for a model with $\sigma_{E}^2=2\mu K^2$
and no $B$. For
comparison we also show what is expected for SCDM with an optical depth
due to reionization $\tau_{ri}=1.0$. Note that even for this large
optical depth the expected signal to noise is very small. The smaller
$k$ modes have smaller S/N in the $\tau_{ri}=1$ model than in the
flat spectra model because large scale polarization power spectra
decreases rapidly  with $l$ in all realistic models. In figure  
\ref{figure4} we also show the $E$ spectra for a model with
$\tau_{ri}=1$. All models where the universe reionizes at an early
enough epoch have a peak in the large angular scale polarization
spectra \cite{zalreio}. 
For this model $l_{peak}\sim 20$ and in general $l_{peak}\sim
2\sqrt{z_{ri}}$ with $z_{ri}$ the redshift at which the universe
reionizes. We can see that the window function of the experiment are
more sensitive to larger angular scale polarization and thus although
the peak of the $\tau_{ri}=1$ model  has an amplitude of $\sim 6\mu K^2$ the 
flat spectrum model with amplitude $2 \mu K^2$ produces a higher S/N
for most $k$. 

The experiment can only detect cosmological polarization for extreme models
with a very early reionization. On the other hand it has the virtue of
being able to put separate constraints on $E$ and $B$ because it
measures both $Q$ and $U$. This is the feature of the experiment that
we are most interested in. We
will analyze simplified models with constant power spectrum
for both $E$ and $B$. This will allow us to study what kind of
constraints a ring experiment can put on $E$ and $B$ separately, not
relying on the different shapes of these two contributions.  
The approximation  also makes visualization of the results easier and
it is also straight forward to get results for any other noise
level, one just needs to rescale the values of $\sigma_{(E,B)}^2$
accordingly.

\section{Estimating Parameters}

A useful way of characterizing the amount of information a given
experiment can provide is through the Fisher matrix. If one is
interested in constraining a set of parameters ${\bf s}$, the Fisher
matrix is defined to be
\begin{eqnarray}
{\bf F}_{ij}={1\over 2} tr[{\bf A}_i{\bf A}_j] \nonumber \\
{\bf A}_i={\bf C}^{-1}{\partial {\bf C} \over \partial s_i}. 
\end{eqnarray}
For example 
the minimum error bars that can be obtained on a given parameter is 
$\Delta s_i=\sqrt{{\bf F}^{-1}_{ii}}$. 

The Fisher matrix can be used to construct minimum variance 
quadratic estimators for the
different parameters \cite{max2,bkj},
\begin{eqnarray}
{\hat s}_j&=&{\bf x}^{T}{\bf D}_j{\bf x}-
tr({\bf N}{\bf D}_j)\nonumber \\
{\bf D}_j&=&{1 \over 2 {\bf F}_{jj}} {\bf C}^{-1}
{\partial {\bf C} \over \partial s_j}{\bf C}^{-1}.
\end{eqnarray}
These estimators also have the virtue of being unbiased, 
$\langle {\hat s}_j \rangle= s_j$. The estimator
depends on the parameters one intends to estimate through the
correlation matrix so some iterative approach may be  needed to implement
this method. This procedure can be seen as an iterative way of
maximizing the likelihood \cite{bkj}. On the other hand a reasonable
choice of underlying spectrum usually leads to a good estimator which has
a slightly higher variance than could be achieved if the ``correct''
underlying model was used to build the estimator. Perhaps a more
important problem is that one also makes a mistake in the error bars
of the estimator which
could lead to errors in the subsequent determination of
cosmological parameters from the data.
We will discuss this
further in the following section.

\subsection{Small scale experiment}

Let us start with the simplest application of these formulas. Let us
imagine that we are only interested in determining one parameter from
the small ring experiment, we will assume that the shape of the
spectrum is given and that we want to obtain an
amplitude. We will parametrize the power spectrum as $C_{El}=\beta
{\bar C}_{El}$, ${\bar C}_{El}$ is a fiducial spectrum which might be
for example SCDM. The correlation matrix and its derivative are then,
\begin{eqnarray}
C(k,k^{\prime})&=&(\beta \alpha_k+w^{-1})\delta_{k,k^\prime}
\nonumber \\
{\partial C(k,k^{\prime}) \over \partial \beta}&=& \alpha_k
\delta_{k,k^\prime}.
\end{eqnarray}
where we have defined $\alpha_k=\sum_{l\ge k}(2l+1)/4\pi W_{1,lk}
{\bar C}_{El}$. Note that $\alpha_k$ is both a function of the
experimental set up, ie. the ring and beam size and of the
fiducial cosmological model.  

\begin{figure}[t]
\vspace*{6cm}
\caption{Relative error in $\beta$ for three different underlying
cosmological models as a function of the size of the ring
$\theta$. The two curves in each case are for beam size $0.22^o$ on the
top and $0.2^o$ on the bottom.}
\includegraphics{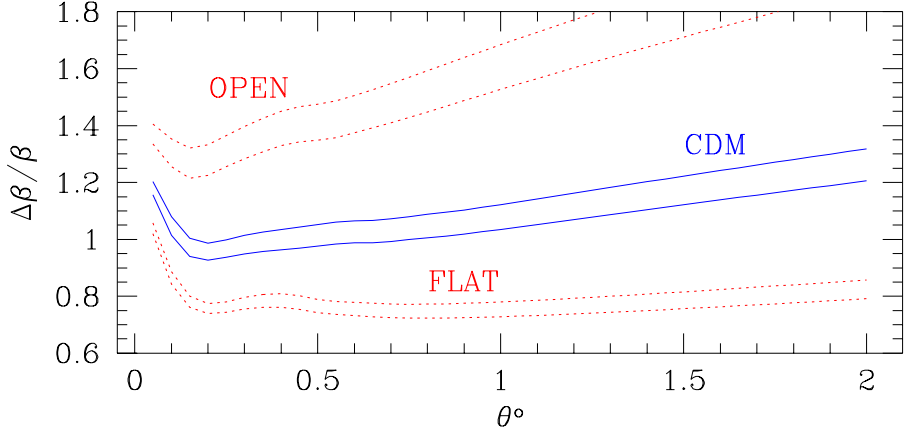}
\label{figure6}
\end{figure}

As we are only estimating one parameter the Fisher matrix is just
a number ($F$),
\begin{equation}
F={1\over 2} \sum_k {\alpha_k^2 \over (\beta \alpha_k+w_P^{-1})^2}. 
\end{equation}
We can use this Fisher ``matrix'' to obtain
the expected error bars on $\beta$,
\begin{eqnarray}
{\Delta \beta \over \beta}&=&{1 \over \beta \sqrt{F}} \nonumber \\
&=& \sqrt{2\over \sum_k 1/(1+w^{-1}_P / \beta \alpha_k)}.
\label{errorprinc}
\end{eqnarray}
This can be interpreted as saying that the effective  number of modes
contributing informations is $N_{eff}= \sum_k 1/(1+w^{-1}_P / \beta
\alpha_k)$, where clearly the modes with a higher signal to noise
contribute more information ($S/N|_k=\beta \alpha_k/w^{-1}_P$). 

\begin{figure}[t]
\vspace*{6cm}
\caption{$E$ polarization power spectra for the three models discussed
in the text.}
\includegraphics{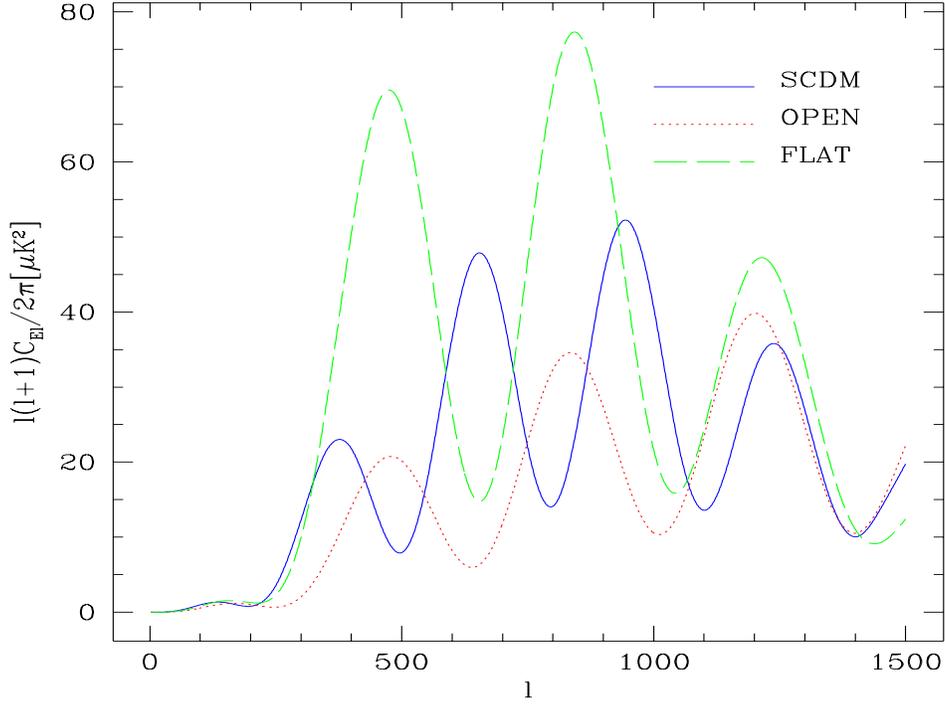}
\label{figure7}
\end{figure}

We can use equation (\ref{errorprinc}) to
investigate the effect of changing the size of the ring
$\theta$, which enters
which enters in the calculation of  
$\alpha_k$. Figure \ref{figure6} shows $\Delta \beta /\beta$ as a
function of $\theta$ for three cosmological models. The $E$
polarization spectra of
these models is shown in figure \ref{figure7}, we choose standard CDM
(SCDM) and two other models that fit well the available temperature
data \cite{models}. One is an open model with 
$\Omega_m=0.85$, $\Omega_b h^2=0.026$,  $h=0.4$ and $n=0.91$.
The other is a flat CDM model with 
$\Omega_b=0.05$ and  $h=0.3$. 
It is interesting to note that while this two
models have similar temperature spectrum up to the first acoustic peak,
within the accuracy
of the current CMB measurements, the amplitude of the polarization
spectra differ significantly. Figure \ref{figure6} shows that this
experiment may be able to detect polarization in the foreseeable
future, at least for some cosmological models. Increasing the
sensitivity by reducing the noise and the angular resolution, thus
going after the larger small scale polarization, will increase the
prospects of the experiment as well.

The quadratic estimator for beta also takes a very simple form,
\begin{equation}
\hat \beta={\sum_k (|{\bf x}_k|^2-w_P^{-1})\alpha_k /(\beta
\alpha_k+w_P^{-1})^2 \over \sum_k 
\alpha_k^2 /(\beta \alpha_k+w_P^{-1})^2},
\end{equation}    
which is nothing more than an inverse variance 
weighing of each mode. 
It is straight forward to check that the ensemble average of
$\hat \beta$ is actually $\beta$
($\langle |{\bf x}_k|^2 \rangle=\beta \alpha_k+w_P^{-1}$),
it is an unbiased estimator.

\begin{figure}[t]
\vspace*{6cm}
\caption{Variance of the ${\hat\beta}$ estimator if the underlying model has
$\beta=1$ but a fiducial model with a different $\beta$ is used to
construct the quadratic estimator. The curves have been scaled using the
minimum variance value, when the $\beta=1$ model is used to
construct the estimator.}  \includegraphics{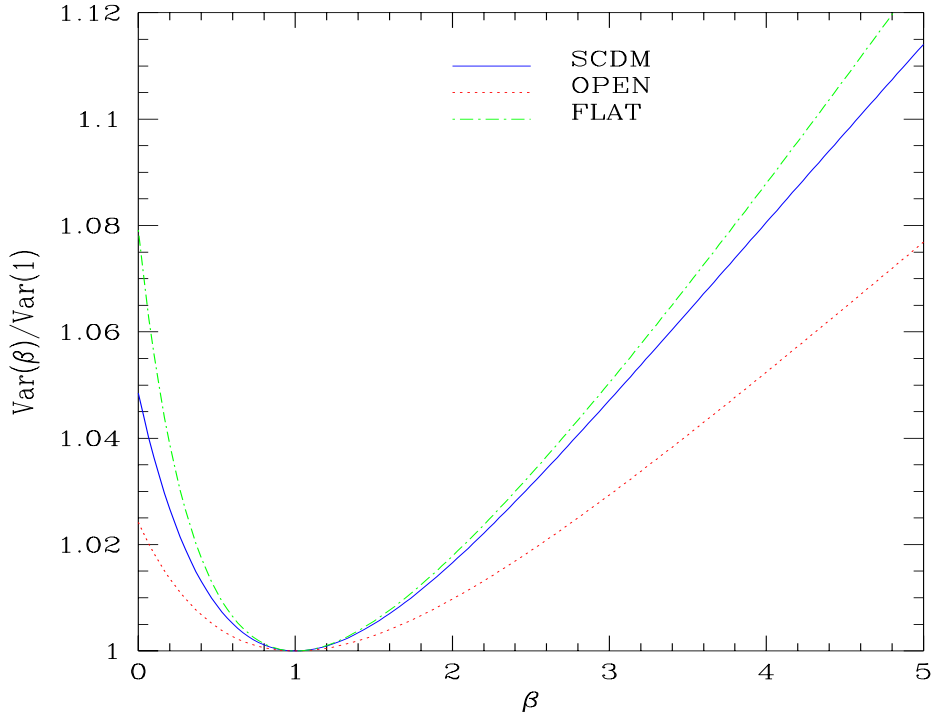}
\label{figure8}
\end{figure}

In this very simple example we can  understand what happens if
we construct the estimator using the ``wrong'' power spectra. 
Let us assume that the
underlying model is SCDM model (ie. that the underlying $\beta=1$ so
that $\langle |{\bf x}_k|^2 \rangle=\alpha_k+w_P^{-1}$) 
but that we use a different $\beta$ to construct the estimator. The
first thing that is easy to check is that $\langle {\hat \beta} \rangle=1$
always, the estimator is always unbiased even if we use the incorrect
$\beta$ to construct it. 
What happens is that the variance of the estimator
increases if use the incorrect fiducial model, the variance 
in the estimator is given by
\begin{equation}
\langle {\hat \beta}^2 \rangle - \langle {\hat \beta} \rangle^2=
2 {\sum_k \alpha_k^2/(\beta \alpha_k+w_P^{-1})^2 \times
(\alpha_k+w_P^{-1})^2/(\beta \alpha_k+w_P^{-1})^2 \over [
\sum_k \alpha_k^2/(\beta \alpha_k+w_P^{-1})^2]^2}
\end{equation} 
This function is shown in figure \ref{figure8}, the variance has a minimum at
$\beta=1$. If one chooses an incorrect  value of $\beta$ the
estimator is still unbiased but is not optimal, one could get more
information out of the data (ie. reduce the error bars on $\beta$).  
Note that the if $\beta \sim 2$ which means we are overestimating the
power in the underlying model by a factor of 2, the variance only
increases by a few percent. Even for $\beta\sim 4$ the increase is
less than 10 \%. The curves in figure \ref{figure8} depend on the signal
to noise ratio of the experiment, for example the open model
curve grows more slowly than the other two. This model has less power
so more of the variance in the estimator comes from the detector noise
and so it is less sensitive to the fiducial model.

Unfortunately life is not so simple, it is not true that all
the models we are interested in testing have a spectrum of the form
$C_{El}=\beta {\bar C}_{El}$ as is obvious from figure \ref{figure7}. The
open and flat models were chosen to fit the existing temperature
data \cite{models}, this means that their first acoustic peak is 
approximately in the same place and is approximately of the same height.
The acoustic oscillations in the photon-baryon fluid are responsible
for  both temperature and polarization peaks so
if the temperature peaks of two models coincide in position, 
the polarization ones  will also coincide. They 
not necessarily have the same 
amplitude, as can be seen in figure \ref{figure7}. 
To get the best results in this type of analysis we should  
choose 
$C_{El}=\beta {\bar C}_{El}$ with ${\bar C}_{El}$ corresponding to one
of the models that fit the existing temperature data. Perhaps
polarization can even help distinguish between some of these
models before smaller scale temperature measurements do. 

\begin{figure}[t]
\vspace*{6cm}
\caption{Comparison between the underlying spectrum and the
``recovered'' one. In the top pannel we show $C^{flat}_{El}$ and
$\langle \beta \rangle C_{El}^{open}$. The bottom pannel shows the
S/N for each $k$ mode for the flat model and the scaled open one.} 
\includegraphics{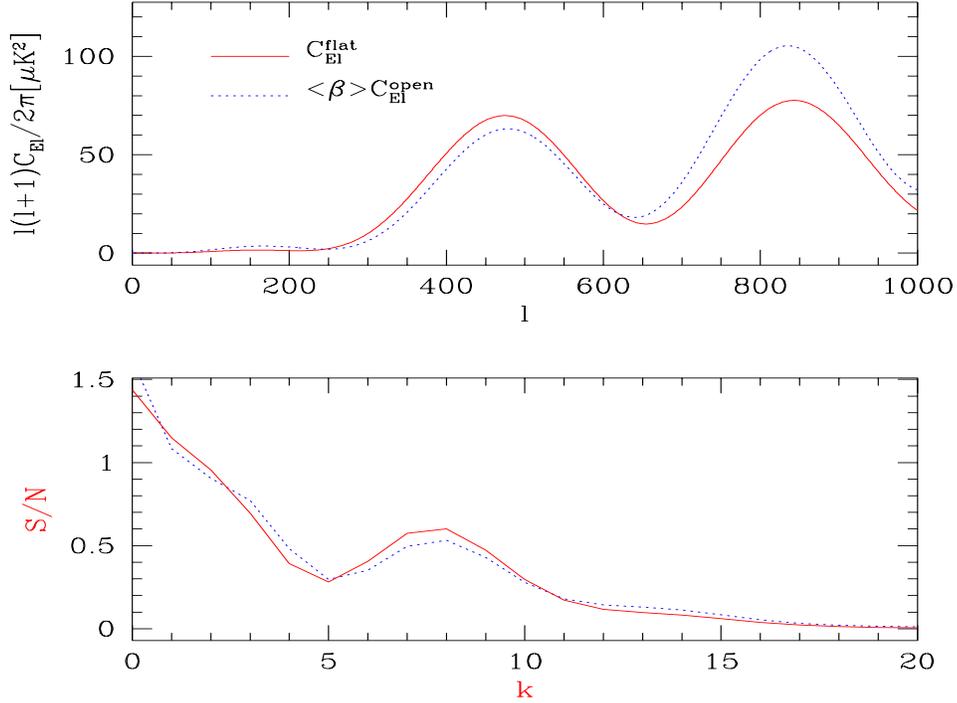}
\label{figure9}
\end{figure}

In fact if we construct the quadratic estimator under the assumption
that the underlying model is the open one, the mean value of
$\beta$ if the real model is the flat one is
$\langle \beta \rangle=3.033$, figure \ref{figure9} compares
$l(l+1)C_{El}^{flat}/2\pi$ and $\langle \beta \rangle
l(l+1)C_{El}^{open}/2\pi$. 
The agreement is very good especially for the S/N curves for the two
models, which is what the experiment is ultimately sensitive to.

In any case $\beta$ should be taken just as an indication of the power
in polarization, and because this experiment is sensitive to such a
wide range in $l$ it is better to assume some realistic spectrum shape
given that the low $l$ polarization is so suppressed. If we do not do
so and just assume a flat spectrum the weighing of the $k$ modes in
$\hat \beta$ will be very bad, leading to an estimator with a very
large variance. Of course one could divide the $l$ range in many bins
so that the flat approximation in each bin is reasonable, but the S/N
in the experiment does not permit that yet. The best way to go is to
assume a shape for the spectrum corresponding to some model that fits
the existing temperature data well.

\subsection{Wisconsin Experiment}

Let us now consider the Wisconsin experiment,
with the assumption that both $E$ and $B$ spectra are constant, the
correlation matrix for each $k$ is
\begin{eqnarray}
\bf{C}_k&=&\langle {\bf x}_k {\bf x}_k^{\dag} \rangle \nonumber \\
&=&{\bf E}_k \sigma_{E}^2 +{\bf B}_k \sigma_{B}^2 +
w_P^{-1}{\bf 1} \nonumber \\
\nonumber \\
{\bf E}_k&=& \left( \begin{array}{cc}
\alpha_{k,1} & \alpha_{k,3} \\
\alpha_{k,3} & \alpha_{k,2}
\end{array} \right) \nonumber \\
\nonumber \\
{\bf B}_k&=& \left( \begin{array}{cc}
\alpha_{k,2} & \alpha_{k,3} \\
\alpha_{k,3} & \alpha_{k,1}
\end{array} \right), 
\label{wisccorr}
\end{eqnarray}
where $\alpha_{k,j}=\sum_{l\ge k}(2l+1)/4\pi W_{j,lk}$ for $j=1,2$ or $3$
and ${\bf 1}$ is the identity matrix.

Note that there is a symmetry between  $(E,B)$ and $(Q,U)$,
that is to say interchanging $E$ and $B$ and $Q$ and $U$ leaves 
equation (\ref{wisccorr})
unchanged. Consider for example the previous signal to noise plot for a
model with $\sigma_{E}^2=2\mu K^2$ and no $B$ (figure
\ref{figure5}).  In the  
``conjugate'' model having $\sigma_{B}^2=2\mu K^2$ and
no $E$, the $Q$ and $U$ signal to noise plots would just be
interchanged. The signal to noise plots for both parameters are different so at
least in principle 
the $E$ and $B$
contributions can be separated. 

\begin{figure}[t]
\vspace*{6cm}
\caption{Contour plots of the $\sigma_{(E,B)}^2$ error bars that could
be obtained by the Wisconsin experiment 
for a grid of underlying models with $E$ and $B$
amplitudes  between $0 \mu K^2$ and $10 \mu K^2$.} 
\includegraphics{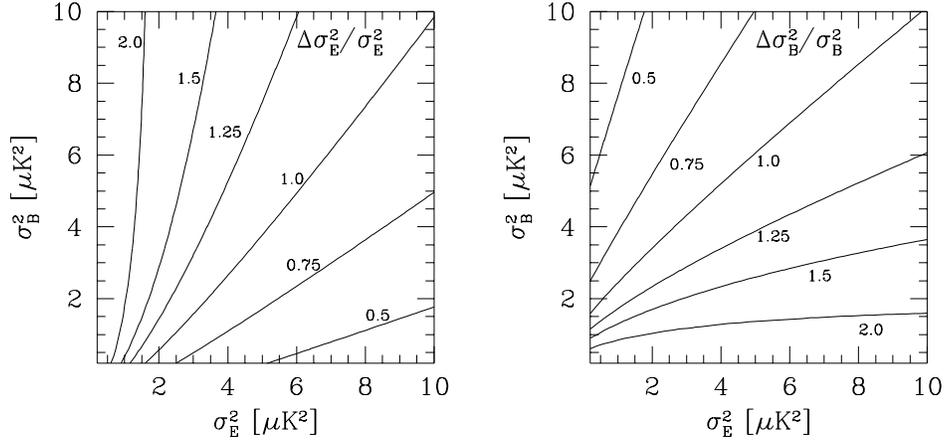}
\label{figure10}
\end{figure}

We have two parameters so
the Fisher matrix is $2\times 2$. 
we used it
to compute the expected error bars on $\sigma_E^2$ and $\sigma_B^2$
in a grid of underlying models. Contour plots of the error bars are
shown in figure \ref{figure10}. One interesting feature of these plots is that
because of the symmetry discussed above both figures are essentially
the same, one can go from one to the other by simply interchanging the
axis. 

\begin{figure}[t]
\vspace*{6cm}
\caption{Contour plots of $68 \%$ and $95 \%$ confidence in the 
$\sigma_E^2-\sigma_B^2$ plane that could be obtained with the
Wisconsin experiment. The axis have been scaled using the amplitudes
of the underlying models. Two different models are considered.} 
\includegraphics{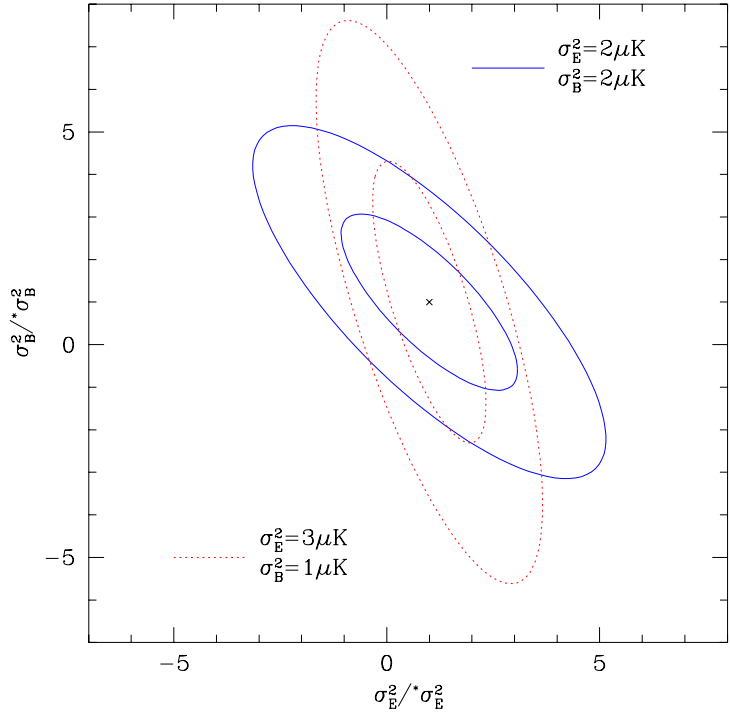}
\label{figure11}
\end{figure}

The other feature that deserves attention is that for a
fixed value of one of the amplitudes, for example $\sigma_E^2$ as one
increases the other one ($\sigma_B^2$) the error bar on $\sigma_E^2$
grows. What happens is that the determination of both amplitudes
is highly correlated, and always the combination that is better
constrained tends to be more aligned with the variable with the
highest amplitude. We show the error ellipses  in the
$\sigma_E^2-\sigma_B^2$ plane for two models in
figure \ref{figure11}. The first model has $\sigma_E^2=2\mu K^2$ and
$\sigma_B^2=2\mu K^2$, the second one $\sigma_E^2=3\mu K^2$ and
$\sigma_E^2=1\mu K^2$. 
Clearly because of the $E-B$ symmetry in the correlation matrix if the
underlying model has $\sigma_E^2=\sigma_B^2$ the sum of both amplitudes
(ie. the total power) is much better constrained than their difference.
When we consider a model with more power in the $E$ channel the
ellipses rotate so that $\sigma_E^2$ is better constrained than $\sigma_B^2$.

\begin{figure}[t]
\vspace*{6cm}
\caption{Expected error bars on $\sigma_E^2$ for models with no $B$ as
a function of the underlying models $\sigma_E^2$. The lower curve
results if we {\it
assume} there is no $B$ rather than fitting for its amplitude.}
\includegraphics{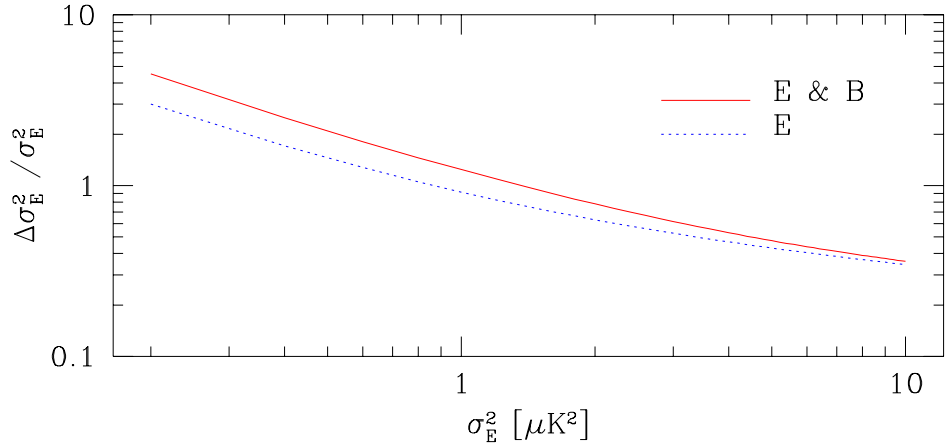}
\label{figure12}
\end{figure}

In figure \ref{figure12} we show the error bars expected for $\sigma_E^2$ under
the assumption that $\sigma_B^2=0$, that is models that lay on the
x-axis of figure \ref{figure10}. We show two curves, one correspond to trying to
determine both $\sigma_E^2$ and $\sigma_B^2$ and the other {\it
assumes} we know  there is no $B$ polarization. This is
actually a good approximation for most models, only defect models that
usually have a large 
vector contribution have a large $B$ contribution. These
models do not seem to be in very good accord with the existing CMB and
LSS data \cite{penurostur}. 

This experiment would be able to detect the cosmological polarization
in an  underlying model with $\sigma_E^2\sim 1 \mu K^2$. 
A model with $\tau_{ri}=1$ could be detected, but this model has
most of its large angular scale power outside the band where the
experiment is most sensitive. The ``reionization peak'' in this case
occurs at $l_{peak}\sim 20$ with an amplitude $6 \mu K^2$, thus a  
smaller beam would greatly increase the sensitivity of the experiment
to the reionization history of the universe, making the experiment
better ``tunned'' for the position of the peak in $l$ space.

It is also clear that a ring experiment can put independent constraints on
both $E$ and $B$ separately if it measures $Q$ and $U$. Although the
assumption that the spectrum is flat across all the band is not very
good in this case, if the noise in the experiment was reduced then
several bins in $l$ could be used and our analysis would become more
realistic.   

\section{Summary and Discussion}

We have studied in detail polarization experiments that measure the
Stokes parameters on a ring in the sky. We have shown that the Fourier
transform of the Stokes parameters on the ring provides natural
variables to analyze the results of the experiment, and have studied
how the correlation matrix of this quantities depend on the $E$ and
$B$ power spectra. We calculated the window functions and discussed
how it depends on the size of the ring. 

Experiments that measure both Stokes parameters are capable of
determining separately the contributions from both $E$ and $B$. 
They can do so without assuming a specific shape for each of the
spectra  because $E$ and $B$ contribute differently to $Q$ and
$U$ autocorrelations.
In fact for  $k=0$ the $Q$ mode only depends on $E$ while the $U$
mode is only sensitive to $B$.

We have calculated the Fisher matrices for these experiments and used
them to estimate their ability to detect the
amplitude of cosmological polarization. We also constructed the
minimum variance quadratic estimator for the parameters of interest
and studied their properties in detail

An experiment targeted at small scale polarization measuring the
Stokes parameters in a small ring with a fraction of a degree resolution, 
may detect cosmological
polarization in the near future. In fact, with sufficient integration it
may help distinguish between models that have very similar temperature
power spectra in the $l$ range measured so far and so could add useful
information.

The Wisconsin experiment is mostly sensitive to large scale
polarization but because the signal in this part of the spectrum is very
small it will be harder for it to detect
polarization. On the other hand 
the measurement of both $Q$ and $U$ will allow the
experiment to put separate constraints on $E$ and $B$.

The prospect of detecting polarization is increasing quickly and
in the near future we may have a positive detection.  We  will 
then be able 
to study very interesting characteristics of our universe, like its ionization
history or the presence of vector and tensor modes. 

\section*{Acknowledgments}
I am very grateful to Uro\v s Seljak for his encouragement 
and many useful discussions. I want to thank J. Gundersen
B. Keating, S. Staggs and
P. Timbie for providing me with information experimental issues. This
work was supported by NASA grant NAG5-2816.

\section {Appendix}

In this appendix
we will consider the window functions for the Wisconsin Experiment. 
This experiment will be pointing in the vertical direction scanning a
ring in the sky with $\theta=43^o$. The instrument will be rotated in
steps of $45^o$ to measure both $Q$ and $U$ \footnote{This is one of
the proposed strategies, a continuous rotation of the instrument is
also being considered.}.
The signal at each pixel is given by
\begin{equation}
X_j={1 \over \delta t} \int_{t_o-\delta t/2}^{t_o+\delta t/2} dt
\int d\Omega^\prime B({\hat n}(t),{\hat
n}^\prime) X(\hat n)
\end{equation}
where $X$ stands for the Stokes parameter being measured, $j$ represents the
pixel and $\delta t$ is the time interval the instrument is measuring each
particular Stokes parameter before being rotate by $45^o$ degrees. 
The beam is represented by the
convolution with $B(({\hat n}(t),{\hat n}^\prime)$ and ${\hat n}(t)$
is the instantaneous position of the beam, in this case given by 
$\theta=43^o$ and $\phi=\omega_s t$ with $\omega_s=2\pi/$day.

We can use the expansion in harmonics (equation
\ref{Pexpansion2}) to calculate the convolution with the beam. For a
gaussian beam this gives the usual factor $\exp{[-l(l+1)\sigma_b^2/2]}$. For
example
\begin{eqnarray}
Q_j&=&{1 \over \delta t}\int dt [-\sum_{lm} 
a_{E,lm} e^{-l(l+1)\sigma_b^2/2} X_{1,lm}(\theta,\phi(t))
\nonumber \\ 
&& +i a_{B,lm} e^{-l(l+1)\sigma_b^2/2} X_{2,lm}(\theta,\phi(t))] 
\end{eqnarray}
and a similar expression applies to $U$. 

The time integration is simple, the dependence is only in the
$\exp{[im\phi(t)]}$ factors inside $X_{(1,2),lm}$. The window function
then acquires another factor from
\begin{equation}
{1 \over \delta t}\int dt e^{im\phi(t)}=e^{im\phi_j} {\sin (m \Delta
\phi/2) \over m \Delta \phi /2 }
\end{equation}  
with $\Delta \phi= \omega_s \delta t$, the size of the pixel. 
The combined window is 
$B_{lm}= \exp{[-l(l+1)\sigma_b^2/2]}{\sin (m \Delta
\phi/2)/ (m \Delta \phi /2) }$. The importance of this last term
depends on the size of the pixel, and can be made irrelevant by
choosing small enough pixels. We ignored this factor in the main text
for simplicity,  but it can trivially be accounted for.

\end{document}